\documentstyle[12pt,epsf]{article}
\font\twlmsy=msym10 at 12pt
\font\sevenmsy=msym8
\font\fivemsy=msym6
\newfam\Bbbfam
\textfont\Bbbfam=\twlmsy
\scriptfont\Bbbfam=\sevenmsy
\scriptscriptfont\Bbbfam=\fivemsy
\def\Bbb{\fam\Bbbfam}

\newcommand{\rf}[1]{(\ref{#1})}

\newcommand{\beq}{\begin{equation}}
\newcommand{\eeq}{\end{equation}}
\newcommand{\bea}{\begin{eqnarray}}
\newcommand{\eea}{\end{eqnarray}}
\newcommand{\beas}{\begin{eqnarray*}}
\newcommand{\eeas}{\end{eqnarray*}}
\newcommand{\beqs}{\begin{displaymath}}
\newcommand{\eeqs}{\end{displaymath}}

\newcommand{\cB}{{\cal B}}

\newcommand{\cT}{{\cal T} }

\newcommand{\ben}{\begin{equation}}
\newcommand{\een}{\end{equation}}

\newcommand{\bdm}{\begin{displaymath}}
\newcommand{\edm}{\end{displaymath}}

\newcommand{\pa}{\partial}

\newcommand{\bbR}{{\Bbb R}}

\newcommand{\baG}{\overline{G}}
\newcommand{\bac}{\bar{\chi}}
\newcommand{\bab}{\bar{\beta}}
\newcommand{\kab}{\bar{\kappa}}
\newcommand{\bam}{\bar{\mu}}
\newcommand{\bs}{{\bf s}}
\newcommand{\boob}{}

\begin{document}
\topmargin 0pt
\oddsidemargin 5mm
\headheight 0pt
\topskip 0mm

\addtolength{\baselineskip}{0.20\baselineskip}

\pagestyle{empty}

\hfill 31st October 1994

\begin{center}

\vspace{18pt}
{\Large \bf The phase diagram of an Ising model on a polymerized random
surface}

\vspace{2 truecm}

{\em Thordur Jonsson\footnote{e-mail: thjons@raunvis.hi.is} }

\bigskip

Raunvisindastofnun Haskolans, University of Iceland \\
Dunhaga 3, 107 Reykjavik \\
Iceland

\bigskip
\bigskip

{\em John F. Wheater\footnote{e-mail: jfw@thphys.ox.ac.uk}}

\bigskip

Department of Physics, University of Oxford \\
1 Keble Road, Oxford OX1 3NP\\
United Kingdom
\vspace{3 truecm}

\end{center}

\noindent
{\bf Abstract.} We construct a random surface model with a
string susceptibility exponent
$\gamma=1/4$ by
taking an Ising model on a random surface and introducing an additional
degree of freedom which amounts to allowing certain
outgrowths on the surfaces.
Fine tuning the Ising temperature and the weight factor for
outgrowths we find a triple point where $\gamma =1/4$.  At this point
magnetized
and nonmagnetized gravity phases meet a branched polymer phase.

\vfill
\noindent Oxford preprint OUTP94/28P
\newpage
\pagestyle{plain}

\section{Introduction} A few years ago  it was discovered that by
allowing surfaces in matrix models to touch at points, associating a
coupling constant to the touching and fine tuning
this coupling the string susceptibility
exponent could jump from $\gamma$ to $\gamma /(\gamma -1)$
\cite{das,gaume,korchemsky}.
It was then shown in \cite{durhuus} working directly with
triangulated random surfaces that this is a generic phenomenon in random
surface theories.  In \cite{adj} a simple random surface
model was studied where the scenarios of \cite{durhuus} were explicitly
realized and one could construct a random surface theory with $\gamma =1/3$
from one with $\gamma =-1/2$.

Here we generalize this construction, starting with the Ising model on a
random surface and introducing an additional degree of freedom with an
associated coupling.  We map the phase diagram of this theory and show that
there are three phases: magnetized and unmagnetized
gravity phases both of which have $\gamma =-1/2$ and a branched polymer phase
with $\gamma =1/2$.  The exponent $\gamma$ takes the value $-1/3$
 on the line where the magnetized and unmagnetized gravity
phases meet
and on the line separating the
branched polymer phase from the gravity phase we have generically $\gamma
=1/3$.
The three phases meet in a triple point where $\gamma =1/4$.

It has been believed for some time that $\gamma >0$ may be associated with
$c>1$ matter fields interacting with 2-dimensional quantum gravity.  This
is one of the reasons \boob why the transmutation of $\gamma$ found in
\cite{das,gaume,korchemsky,durhuus} is interesting.  In a recent paper
\cite{klebanov} it
is argued on the basis of Liouville theory calculations that the jump of
$\gamma$ to a positive value
is not related to $c>1$ but rather to a different solution of
a $c<1$ theory.  From the point of view of discretized models it is not
clear that this is the correct interpretation.  Studies of models that
manifestly have $c>1$, e.g. many Ising or Potts models on a random surface,
indicate that $\gamma$ depends only on $c$ \cite{amb} and the behaviour of such
models is analogous to the ones with a transmuted $\gamma >0$.

\section{Definition of the model}
Let $\cT$ denote the collection of all triangulations of the
disc where the boundary is just one link.  If $T\in\cT$ we allow two triangles
in $T$ to share two links so in particular this ensemble includes tadpole
surfaces.  The dual graphs corresponding to surfaces in $\cT$ are all
planar $\varphi ^3$-diagrams with one external leg.
To each triangle $i$ in $T$ we associate an Ising spin variable $\sigma _i$.

Before defining our model let us recall some facts about the Ising model
on a random surface.  The one-loop function of the
Ising model on a random surface is given by
\beq
\baG (\mu ,\beta )=\sum _{T\in\cT}e^{-\mu |T|}\sum _{\{\sigma \} }
e^{\beta\sum _{(ij)}(\sigma _i\sigma _j-1)/2},
\eeq
where the sum on $\{\sigma\}$ is over all spin configurations on the
triangulation $T$ and the sum on $(ij)$ is over all nearest neighbour pairs of
triangles in $T$.  Here $|T|$ usually denotes the number of triangles but for
later
convenience we shall let $|T|$ denote the number of interior links in the
triangulation $T$.  Since the number of interior links is linearly related
to the number of triangles the  values of critical exponents do not depend
upon this change.  The susceptibility of the Ising model is defined by
\beq
\bac (\mu ,\beta )=-{\pa\over\pa\mu }\baG (\mu ,\beta ) .
\eeq
This theory was solved exactly in
\cite{kazakov,boulatovkazakov} and the critical exponents calculated.
We shall make use of the following facts from \cite{kazakov,boulatovkazakov}:
For each $\beta \geq 0$ there is $\mu _c^I(\beta )>0$ such that $\bac
(\mu ,\beta )$ is analytic in $\mu$ for $\mu >\mu _c^I(\beta )$ and infinite
for  $\mu <\mu _c^I(\beta )$.  As $\mu\downarrow  \mu _c^I(\beta )$
\beq
\bac (\mu ,\beta )\sim (\mu - \mu _c^I(\beta ))^{-\gamma ^I(\beta )}
+\mbox{\rm less singular terms}
\eeq
where
\beq
\gamma ^I(\beta )=\left\{\begin{array}{ll}-1/2,&\beta\neq \beta _c^I\\
-1/3,&\beta =\beta _c^I\end{array}\right.\label{ising}
\eeq
and $\beta _c^I= -{1\over 2}\log\left( {1\over 27}(2\sqrt{7}-1)\right)$.

The model we wish to study is an Ising model on a random surface
as described above with an additional
degree of freedom which we shall call {\em outgrowths}.
If $\ell$ is an interior link in a triangulation
$T$, we put an outgrowth on the triangulation
at $\ell$ by cutting it open along $\ell$, gluing two sides of a new
triangle to the boundary of the cut and attaching a surface in $\cT$
to the remaining boundary link in the new triangle.  We associate a
\boob non-negative weight
factor $\lambda$ to each outgrowth. The
extra triangle used for gluing an outgrowth on the underlying
surface $T$ carries an Ising spin
which interacts with its neighbours, see Fig. 1.
The triangles and links in the
outgrowth have the same degrees of freedom as
the ones in the underlying surface.
One can think of the outgrowths as being defined by the phase boundaries
of an additional
restricted Ising spin system on the surface where phase boundaries
in the restricted system are only allowed to have length 1, cf. \cite{adj}.

Another way to think of
the model is the following:  For each triangulation we consider all
interior loops of length 1.  To each such loop we assign a variable with
two values, blue and red say.  If the colour is red this loop is the
boundary of an outgrowth and there is a corresponding weight factor $\lambda$
associated with it.  If the colour is blue the loop is not the boundary of an
outgrowth and no multiplicative factor is associated to it.
In the partition function we then sum over all colour
assignments to the loops of length 1.

Let $G(\mu ,\beta , \lambda )$ denote the one-loop function in our theory.
If the value of the boundary spin is fixed to be $\sigma$ we denote
the one-loop function by
$G_{\sigma}(\mu ,\beta ,\lambda )$.
In the absence of an external magnetic field, as will be the case in this
paper,
$G_{\sigma }$ is independent of $\sigma$ and $G=2G_{\sigma}$.
We can express the one-loop function as
\bea
G(\mu ,\beta ,\lambda )&=&\sum _{T\in\cT}e^{-\mu |T|}\sum _{\{\sigma \} }\prod
_{\ell\in T}\nonumber\\
& &\left( e^{\beta (\sigma _i\sigma _j -1)/2}+\lambda \sum _{\sigma ,
\sigma '}e^{\beta (\sigma _i\sigma +\sigma _j\sigma +\sigma \sigma '-3)/2}
e^{-2\mu}G_{\sigma '}(\mu ,\beta ,\lambda )\right),\label{5}
\eea
where $i$ and $j$ are the triangles next to the link $\ell$ in the
triangulation $T$, $\sigma$ is the spin variable associated to the intermediate
triangle for an outgrowth and $\sigma '$ is the value of the boundary spin
of the outgrowth, see Fig. 1.
The first term in the \boob parenthesis on the right side of \rf{5}
corresponds to the case when there is no outgrowth on the link $\ell$
and the second term corresponds to the presence of an arbitrary outgrowth.

Summing over the spin variables $\sigma$ and
$\sigma '$ the one-loop function becomes
\bea
 G(\mu ,\beta ,\lambda )&=&\sum _{T\in\cT}e^{-\mu |T|}\sum _{\{\sigma \} }\prod
_{\ell\in T}\nonumber\\
& &
\left( e^{\beta (\sigma _i\sigma _j -1)/2}+ 2\lambda
e^{-2\mu}e^{-3\beta /2}\cosh ^3{\beta\over 2}
(1+\sigma _i\sigma _j\tanh ^2{\beta\over 2})G\right).
\eea
The function $G$ can be rewritten as the one-loop function
of a single Ising model on a random surface with renormalized couplings
$\bam$ and $\bab$, i.e.
\beq
G(\mu ,\beta ,\lambda )=\baG (\bam ,\bab )\label{identity}
\eeq
 provided
\beq
e^{-\bam}e^{\bab (\sigma _i\sigma _j -1)/2}=
e^{-\mu}
\left( e^{\beta (\sigma _i\sigma _j -1)/2}+ 2\lambda
e^{-2\mu}e^{-3\beta /2}\cosh ^3{\beta\over 2}
(1+\sigma _i\sigma _j\tanh ^2{\beta\over 2})G\right).
\eeq
Since $\sigma _i\sigma _j=\pm 1$ we have \boob two equations for \boob two
unknowns which
are readily solved and we find\boob
\beq
\bam =\mu -\log\left( 1+{\lambda\over 2}e^{-2\mu }(1+e^{-\beta})(1+e^{-2\beta})
G(\mu ,\beta ,\lambda )\right)\label{x}
\eeq
\beq
\bab =-\log{e^{-\beta} +\lambda e^{-2\mu}e^{-\beta} (1+e^{-\beta })G(\mu ,\beta
,\lambda )\over 1+{\lambda\over 2}e^{-2\mu }(1+e^{-\beta})(1+e^{-2\beta})G(
\mu ,\beta ,\lambda )}.\label{y}
\eeq

If the couplings $\beta$ and $\lambda$ are fixed there is a critical value of
$\mu$ which we denote by $\mu _c(\beta ,\lambda )$ such that the one loop
function $G$ is analytic in $\mu $
for $\mu >\mu _c(\beta ,\lambda )$ and infinite for $\mu <\mu _c (\beta ,
\lambda )$.
In fact one can show that the set $\cB =\{ (\mu ,\beta ,\log\lambda ):
G(\mu ,\beta ,\lambda )<\infty\}$ is a convex
subset of $\bbR ^3$ and $G$ is a real analytic function in its interior.
Note that the renormalized couplings $\bam$ and $\bab$ are
functions of all the unrenormalized couplings $\mu$, $\beta$ and $\lambda$.
Due to \rf{identity} we clearly have
\beq
\bam (\mu _c(\beta ,\lambda ),\beta ,\lambda )\geq \mu _c^I(\bab
(\mu _c(\beta ,\lambda) ,\beta ,\lambda )).
\eeq

\section{The phase diagram}
Here we derive a relation between the susceptibility of our model
\beq
\chi (\mu ,\beta ,\lambda )=-{\pa \over \pa\mu}G(\mu ,\beta ,\lambda )
\eeq
and the susceptibility of the random surface
Ising model with renormalized couplings.
This will enable us to calculate the critical exponent $\gamma (\beta ,
\lambda )$ defined by
\beq
\chi (\mu ,\beta ,\lambda )\sim (\mu -\mu _c(\beta ,\lambda ))^{-\gamma
(\beta ,\lambda )}+\mbox{\rm less singular terms}
\eeq
as $\mu\downarrow\mu _c(\beta ,\lambda )$.

First note that
\beq
\chi =\bac{\pa\bam\over\pa\mu}-{\pa\baG\over\pa\bab}{\pa\bab\over\pa\mu}
\label{diff}.
\eeq
If we define the function $C(\bam ,\bab )$ by the equation
\beq
{\pa\baG\over\pa\bab}=\bac (\bam ,\bab )C(\bam ,\bab ),\label{defc}
\eeq
then it easy to see that $-1/2 \leq C\leq 0$.  In order to simplify
some of the formulas in the sequel we put
\beq
\Lambda ={\lambda\over 2}e^{-2\mu}(1+e^{-\beta})(1+e^{-2\beta}).
\eeq
After a calculation, using \rf{identity}, \boob\rf{x}, \rf{y} and \rf{diff}, we
find
\beq
\chi (\mu ,\beta ,\lambda )={N(\mu ,\beta ,\lambda )\bac (\bam ,\bab )\over
D(\mu ,\beta ,\lambda )}\label{z},
\eeq
where\boob
\beq
N(\mu ,\beta ,\lambda )
=1+{2\Lambda \baG\over 1+\Lambda \baG}-
{2\Lambda C\baG\tanh \beta\over
\left(1+\Lambda \baG {2\over 1+e^{-2\beta}}\right)
(1+\Lambda \baG)}
\eeq
and
\beq
D(\mu ,\beta ,\lambda )=1-{\Lambda
\bac\over 1+\Lambda \baG} +
{\Lambda C\bac \tanh \beta\over
\left(1+\Lambda \baG {2\over 1+e^{-2\beta}}\right)
(1+\Lambda \baG)}.\label{19}
\eeq

Note that $N$ is a positive uniformly bounded function.
Let us fix $\beta$ and $\lambda$ and take $\mu$ larger than its critical
value $\mu _c(\beta ,\lambda )$.
As we lower $\mu$,
a singularity is eventually encountered
in $\chi (\mu ,\beta ,\lambda
)$
for one or both of
two reasons:  Either $\bam$ reaches the critical value
of the cosmological constant of the random surface Ising model, $\mu _c^I$, or
$D$ becomes zero.
If $\lambda $
is small enough then the denominator $D$
is positive for any $\mu$ and $\beta$ because $\baG$ and $\bac$ are
bounded functions.
On the other hand, the last term in \rf{19} is negative definite so $D$ is
negative if
\beq
\Lambda\bac\geq 1+\Lambda\baG .\label{20}
\eeq
Since $\bac\geq 2\baG$ for all values of the coupling constants, the
inequality \rf{20} holds if $\Lambda\baG\geq 1$.
We shall indeed prove that there is a critical line separating the
region where $D=0$ at the critical point from a region where $D >0$ at the
critical point.

It will be convenient in the remainder of this paper to regard $D$, $N$,
$\mu _c$ etc. as functions of $\Lambda$ rather than $\lambda$.  This
amounts to a smooth change of coordinates in the coupling constant space.
We claim the following: For any value of $\beta$
there is a value of $\Lambda$ which we denote by $\Lambda _c(\beta )$ such that
\beq
D(\mu ,\beta ,\Lambda )>0
\eeq
 for all
$\mu\geq\mu _c(\beta ,\Lambda )$ provided $\Lambda <\Lambda _c(\beta )$ and
\beq
D(\mu _c(\beta ,\Lambda ),\beta ,\Lambda )=0
\eeq
for $\Lambda \geq\Lambda _c(\beta )$.
Furthermore,
\beq
\bam (\mu _c(\beta ,\Lambda ),
\beta ,\Lambda )>\mu _c^I(\bab (\mu _c(\beta ,\Lambda ),\beta ,\Lambda ) )
\eeq
in the region $\Lambda >\Lambda _c(\beta )$.

In order to prove the claim we consider lines in the coupling constant
space with fixed values of $\bam$ and $\bab$.  These lines can be
parametrized by $\Lambda$
and they constitute a
fibration of the coupling constant space.  First we observe, using\boob
\beq
e^{-\bab}=e^{-\beta}{1+{2\Lambda\baG\over 1+e^{-2\beta}}\over
1+\Lambda\baG},\label{nice}
\eeq
that on each such line $\beta$ is
an increasing smooth function of $\Lambda$, $\beta =\bab$ at $\Lambda=0$ and
\boob $\beta =\bab +\log(1+\sqrt{1-e^{-2\bab}})$ at $\Lambda =\infty$.
In order to prove our claim it therefore suffices to show that $D$ has
exactly one zero on each line where $\bab$ and $\bam$ are fixed.  On such a
line
the function $C$, defined in \rf{defc}, is constant
and in view of \rf{nice} we can write
\beq
D=1-{\Lambda \bac\over 1+\Lambda \baG} +{\Lambda C\bac e^{\bab} e^{-\beta}
\tanh \beta\over (1+\Lambda \baG )^2}
\eeq
when $D$ is restricted to the line.
We saw above that $D<0$ if $\Lambda \baG>1$ so in order to prove our claim
it suffices to show that
\beq
{dD\over d\Lambda} <0
\eeq
for $\Lambda \baG\leq 1$. We find
\bea
{dD\over d\Lambda}&=&{-\bac\over (1+\Lambda \baG )^2}+
C\bac e^{\bab} e^{-\beta }\tanh\beta {d\over d\Lambda}\left({\Lambda\over
(1+\Lambda\baG )^2}\right)\\ && +{\Lambda
C\bac e^{\bab}\over(1+\Lambda\baG )^2}
{d\beta\over d\Lambda} {d\over d\beta}\left(e^{-\beta}\tanh\beta\right).
\eea
The second term on the right hand side above is negative definite
if $\Lambda\baG \leq 1$.  Using
\beq
{d\beta \over d\Lambda}={\baG -e^{-\bab}\cosh\beta\baG\over
e^{-2\beta}+(1+\Lambda\baG)e^{-\bab}\sinh\beta}
\eeq
and the inequality
\beq
1\leq e^{-\bab+\beta}\leq{3\over 2},
\eeq
\boob which follows from \rf{nice} if  $\Lambda\baG\leq 1$, one can now check
by an explicit calculation
that the sum of the two
remaining terms is negative.
This completes the proof of the claim.

In the region $\Lambda > \Lambda _c(\beta)$ the functions
 $\baG$ and $\bac$ are
analytic functions of their arguments as we reach the critical point
and we can calculate the critical exponent
$\gamma (\beta ,\Lambda )$ by the same method as in \cite{durhuus,adj}
and find the generic branched polymer
value $\gamma =1/2$.\boob  In this case
the surfaces are in the branched polymer phase,
as expected for large $\Lambda$,
and the entropy is dominated by outgrowths.  We shall
call the line $\Lambda =\Lambda _c(\beta)$ the $\Lambda$-line.
It is easily seen from the implicit function theorem that $\Lambda _c$ is
a smooth function of $\beta$.

Let us now consider the region $\Lambda <\Lambda _c(\beta)$.  In this case the
denominator $D$ vanishes nowhere and by arguments
analogous to those in \cite{durhuus,adj} we obtain
\beq
\gamma (\beta ,\Lambda )=\gamma ^I
(\bab (\mu _c(\beta ,\Lambda ), \beta ,\Lambda
) ).
\eeq
\boob This is the gravity phase of the model which, as expected for small
$\Lambda$,
 is characterized by
few outgrowths.
  It is easy to check from \rf{x} and \rf{y} that the equations
\beq
\bab (\mu _c(\beta ,\Lambda ), \beta ,\Lambda )=\beta _c^I
\eeq
and
\beq
\bam (\mu _c(\beta ,\Lambda ),\beta ,\Lambda )=\mu _c^I(\beta _c^I)
\eeq
have a unique solution $\beta _c(\Lambda )$
in this region for any value of $\Lambda$.  We shall
call this line the $\beta$-line.  This line separates the
gravity phase of the model into two parts, a magnetized one for
$\beta >\beta _c(\Lambda )$
and an unmagnetized one for $\beta <\beta _c(\Lambda )$, see Fig. 2.
In both these phases $\gamma =-1/2$ but on the $\beta$-line $\gamma =-1/3$
by \rf{ising}.

\boob The $\beta$-line meets the $\Lambda$-line at a unique point;
this is where   the magnetized and unmagnetized
gravity phases meet the branched polymer phase and is therefore
a triple point of the theory.  On the $\Lambda$-line
the numerator and denominator
in \rf{z} are both singular and conspire to realize the scenario of
\cite{durhuus}.  We
can calculate the value of $\gamma$ by the same method as in \cite{adj} and
find $\gamma =1/3$ except at the triple point where $\gamma =1/4$.

\section{Discussion}
We have given an explicit construction of a random surface theory
with $\gamma =1/4$ at one particular point in the phase diagram.
\boob It is interesting, as pointed out in \cite{durhuus},
that this value of $\gamma$ has been seen
in simulations of a random surface in a three dimensional hypercubic
lattice with a weak self-avoidance condition \cite{baumann}.\boob  We have,
however,
not been able to see any relation between that model and the one studied here.


These results can be extended to other spin systems.  Suppose  that,
instead of the Ising spin  we have a vector of spins \bs  ($\bs.\bs=1$)
 at each site and we make the replacement
\beq  e^{{\beta}(\bs_1.\bs_2-1)/2}
 \to 1+\kappa \,\bs_1.\bs_2,\quad 0\le\kappa\le1.
\eeq
 (which in the Ising case is exact up to
a factor depending only on $\beta$).  This leads to the equations
\beq
G(\mu ,\kappa ,\lambda )=\baG (\bam ,\kab )
\eeq
\beq
\bam =\mu -\log\left( 1+{\lambda}e^{-2\mu }
G(\mu ,\kappa ,\lambda )\right)
\eeq
\beq
\kab =\kappa {1 +\lambda e^{-2\mu} \kappa\, G(\mu ,\kappa
,\lambda )\over 1+{\lambda}e^{-2\mu }\,G(
\mu ,\kappa ,\lambda )}.
\eeq
which have the same structure as \rf{identity}, \rf{x}, \rf{y}.
  Random surface models
with these generalized matter fields have not been explicitly solved
so  $\baG (\bam ,\kab )$ is not known; however we see that such models
which belong to the minimal series with $\bar{\gamma}=-1/n$  with
$n=2,3,4,\ldots$  must also realize the scenario of  \cite{durhuus}
  and have a
triple point with  $\gamma=1/(n+1)$  when outgrowths are included.

Instead of using surfaces with boundaries of length 1 and allowing
tadpoles one could define a model analogous to the one constructed here
using surfaces with boundaries consisting of two links and not allowing
tadpoles.  In this case one has to work with two different one-loop functions,
i.e. the one where the boundary spins are aligned and another one where they
point in opposite directions.  It turns out that in order to verify the
existence of a triple point one needs rather delicate estimates on the ratio
of these two one-loop
functions.  There is however little doubt that this model
should have the same critical behaviour as the one we have studied here.

\bigskip
\noindent
{\bf Acknowledgement.} T. J. has benefitted from discussions with
J. Ambj\o rn and B. Durhuus and would like to acknowledge
hospitality at Institut Mittag-Leffler.

\newpage
\noindent
{\bf Figure Caption.}

\bigskip
\noindent
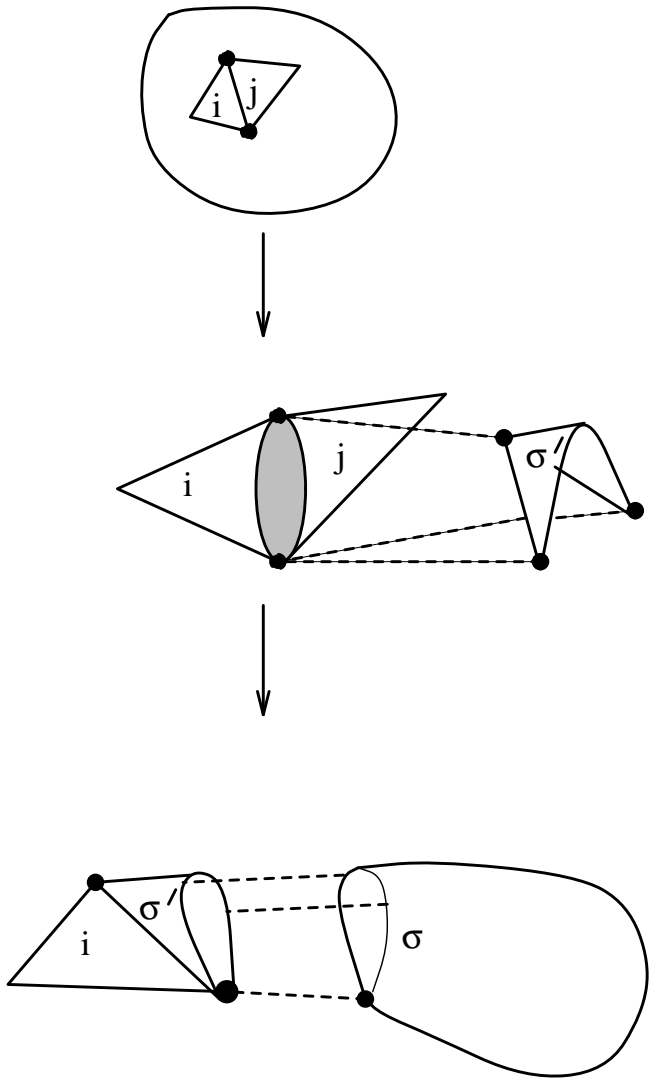
{\bf Fig. 1.} The Figure illustrates the operation of placing an outgrowth
on the link joining the triangles $i$ and $j$.  Here $\sigma$ is the spin on
the intermediate triangle and $\sigma '$ is the boundary spin of the outgrowth.

\bigskip
\noindent
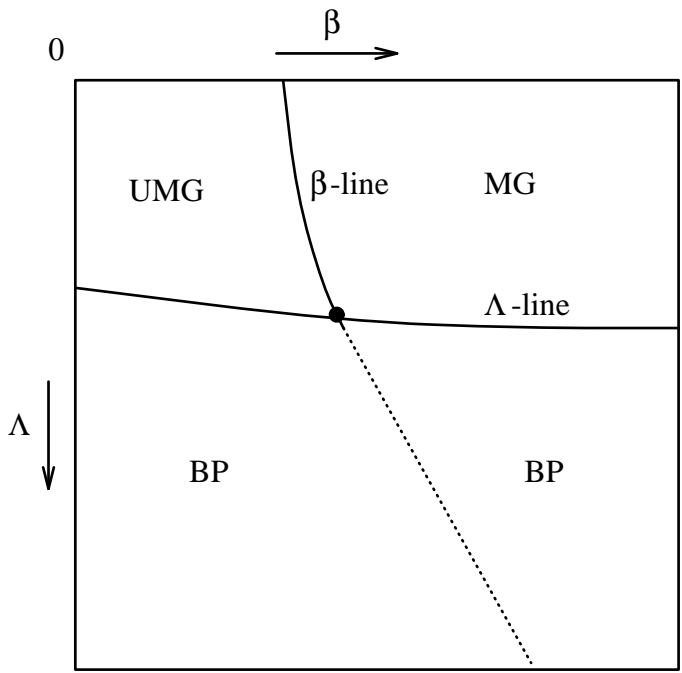
{\bf Fig. 2.} The phase diagram of the model compactified to fill a square.
For $\Lambda >\Lambda _c$ the critical behaviour is governed by
the vanishing of $D$ while $\bam$ and $\bab$
are analytic so the $\beta$-line is not a line of phase transition in
this region.  The different phases are denoted  by UMG (unmagnetized
gravity), MG (magnetized gravity) and BP (branched polymer).

\newpage
\newlength{\figheight}
\setlength{\figheight}{6 truein}
\epsfysize=\figheight
\epsfbox{figure1.eps}
\newpage
\setlength{\figheight}{4 truein}
\epsfysize=\figheight

\epsfbox{figure2.eps}


\begin{thebibliography}{99}
\bibitem{das}S. R. Das, A. Dhar, A. M. Sengupta and S. R. Wadia,
Mod. Phys. Lett. A5 (1990) 1041.
\bibitem{gaume}L. Alvarez-Gaume, J. L. F. Barbon, and C. Crnkovic, Nucl. Phys.
B 394 (1993) 383.
\bibitem{korchemsky}G. P. Korchemsky, Phys. Lett. B 296 (1992) 323;
Mod. Phys. Lett. A7 (1992) 3081.

\boob
\bibitem{durhuus}B. Durhuus, Nucl. Phys. B 426 (1994) 203.

\boob
\bibitem{adj}J. Ambj\o rn, B. Durhuus and T. Jonsson, Mod. Phys. Lett. A9
(1994)
1221.
\bibitem{klebanov}I. R. Klebanov, {\em Touching random surfaces and
Liouville gravity}, PUPT-1486, hep-th/9407167.

\boob
\bibitem{amb}J. Ambj\o rn and G. Thorleifsson, Phys. Lett. B 323 (1994) 7.
\bibitem{kazakov} V. A. Kazakov, Phys. Lett. A 119 (1986) 140.
\bibitem{boulatovkazakov}D. V. Boulatov and V. A. Kazakov, Phys. Lett. B
(1987) 379.
\bibitem{baumann}B. Baumann and B. Berg, Phys. Lett. B 164 (1985) 131.

\end{thebibliography}
\end{document}